\documentclass[authoryear,12pt]{article}
\usepackage[english]{babel}
\usepackage{geometry,float}
\usepackage{amssymb}
\usepackage{amstext}
\usepackage{bm}
\usepackage{graphics}
\usepackage{graphicx}
\usepackage{amsthm,amsfonts,epic,latexsym}
\usepackage{amsmath}
\usepackage{latexsym}
\usepackage{rotate}
\usepackage{lscape}
\usepackage{color}
\usepackage{epsfig,epsf,psfrag}
\usepackage{sectsty}
\usepackage{graphics}
\usepackage{epsfig}
\usepackage{url,hyperref}
\usepackage{multirow}
\usepackage{graphicx}
\usepackage{sectsty}
\usepackage{epsfig,natbib}
\usepackage{rotate}
\usepackage{lscape}
\usepackage{graphicx}
\usepackage{enumitem}
\usepackage{setspace}
\usepackage{anysize}
\marginsize{1in}{1in}{1in}{1in}

\newcommand{\R}{\ensuremath{\mathbb{R}}}
\newcommand{\I}{\mathcal{I}}

\newtheorem{Corollary}{{\scshape Corollary}}
\newtheorem{Remark}{{\scshape Remark}}
\newtheorem{Proposition}{{\scshape Proposition}}

\begin{document}

\title{Shape Outlier Detection and Visualization for Functional Data: the Outliergram}

\author{ANA ARRIBAS-GIL$^\ast$, JUAN ROMO\\[4pt]
\textit{Departamento de Estad\'istica, Universidad Carlos III de Madrid, Getafe, Spain.}
\\[2pt]
{ana.arribas@uc3m.es}}

\markboth%
{A. Arribas-Gil and J. Romo}
{Shape Outlier Detection for Functional Data}

\maketitle

\footnotetext{To whom correspondence should be addressed.}

\begin{abstract}
{We propose a new method to visualize and detect shape outliers in samples of curves. In functional data analysis we observe curves defined over a given real interval and shape outliers may be defined as  those curves that exhibit a different shape from the rest of the sample. Whereas magnitude outliers, that is, curves that lie outside the range of the majority of the data, are in general easy to identify, shape outliers are often masked among the rest of the curves and thus difficult to detect. In this article we exploit the relation between two measures of depth for functional data to help visualizing curves in terms of shape and to develop an algorithm for shape outlier detection. We illustrate the use of the visualization tool, the outliergram, through several examples and assess the performance of the algorithm on a simulation study. Finally, we apply our method to identify outliers in real data sets of growth and mortality curves.}
{Depth for functional data; Outlier visualization; Robust estimation.}
\end{abstract}
\newpage

\section{Introduction}
In many biomedical and public health studies, individual observations are real functions of time, observed at discrete time points. Each curve provides the evolution over time of a certain process of interest for a given individual. When the grid of points is dense enough, and the underlying process is known to be continuous, curves can be treated as functional data. In this context, data can be seen as a sample of curves, in which it is common to observe outlying trajectories. Examples include human growth curves \citep[see, e. g.,][]{RamSil}, time-course microarray experiments \citep{TangMuller}, mortality and fertility rates \citep{HU}, pollutants concentration across time \citep{FebGalMant08} or arterial oxygen saturation \citep{oxy} among others.

As in the univariate or the multivariate case, the presence of atypical observations may affect the statistical analysis of the data. Thus, any preliminary analysis should include an outlier detection step. The challenge in the functional framework is that outlying observations might be very difficult to identify visually. Indeed, an outlier trajectory is not only one that contains atypically high or low values, but also a trajectory that even containing average levels across the whole observation interval may present a different shape or pattern than the rest of the curves of the sample. Following \cite{rainbow}, we will refer to the first type of atypical curves as magnitude outliers and to the second one as shape outliers. Whereas magnitude outliers might be easily detected even with a simple visual inspection of the data, shape outliers will be \emph{hidden} in the middle of the sample and its identification will not be straightforward. That is, a shape outlier is not an observation that lies far away from the sample in terms of the Euclidean distance. Then, measures that can deal with shape features need to be considered in order to be able to distinguish them from the rest of the curves. In this article we combine two sources of information about curve shape given by two measures of depth for functional data, the modified band depth \citep{BandDepth} and the modified epigraph index \citep{HalfRegion11}. Each one provides an ordering of the sample according to different shape features and the relation between them sheds light on shape variation across the sample.

Different methods for outlier detection in functional data have been developed during the last years. Among them several rely on different notions of functional depth \citep{FebGalMant08,SunGenton11, Gervini12,HalfBP}, on robust principal components \citep{rainbow} or on random projections of infinite dimensional data into $\R$ \citep{ FraimanSvarc13}. Also, some distributional approaches have been considered \citep{Gervini09}. While some of these methods may only be sensitive to magnitude outliers, most of them work efficiently at detecting different kinds of outlying trajectories. However, the mechanisms they rely on do not always allow to provide an easy interpretation on why an observation is considered an outlier. In this sense, the objective of this article is two-fold. On the one hand, we propose an algorithm for shape outlier detection. On the other hand, we give a visualization tool that helps understanding shape variation across the sample and provides additional information that can be used to correct the output of the algorithm.

The rest of the article is as follows.  In Section \ref{sdepth} we review the definitions of modified band depth and modified epigraph index and investigate their relationship on a sample of curves. Based on it, in Section \ref{salgo} we propose a visualization tool that helps identifying curves with different shape and propose a rule for shape outliers detection. We evaluate our method and previous existing techniques for outlier detection in functional data through an extensive simulation study in Section \ref{ssim}. In Section \ref{sapp} we apply our method to the Berkeley growth data set and to a set of Australian male log-mortality rates (1901-2003) and compare the results with existing methods. Finally, we conclude the article with a discussion in Section \ref{sdisc}.

\section{Modified Band Depth and Modified Epigraph Index}\label{sdepth}

Let us introduce the modified band depth (MBD) and modified epigraph index (MEI), as firstly defined in \cite{BandDepth} and \cite{HalfRegion11}, respectively. Both measures provide an idea of how central or deep a curve is with respect to a sample of curves. Let $x_1,\dots, x_n$ be $n$ continuous functions defined on a given closed real interval  $\I$. For any $x\in \{x_1,\dots,x_n\}$ we define its modified band depth as
$$MBD_{\{x_1,\dots,x_n\}} (x) =\binom{n}{2}^{-1} \sum_{i=1}^n \sum_{j=i+1}^n \dfrac{\lambda\left(\left\{t\in {\cal I} \,\middle|\, \min( x_i(t),x_j(t)) \leq x(t) \leq \max( x_i(t),x_j(t)) \right\}\right)}{\lambda(\I)},$$
where $\lambda(\cdot)$ stands for the Lebesgue measure on $\R$. If for each pair of curves $x_i$ and $x_j$ in the sample we consider the band that they define in $\I\times\R$ as $\left\{(t,y)\,\middle|\,t\in \I, \, \, \min( x_i(t),x_j(t)) \leq y \leq \max( x_i(t),x_j(t))\right\}$, then $MBD_{\{x_1,\dots,x_n\}} (x)$ represents the mean over all possible bands of the proportion of time that $x(t)$ spends inside a band. The modified band depth is an extension of the original band depth that accounts for the proportion of bands in which a curve is entirely contained \citep[see][for details]{BandDepth}.

The modified epigraph index of  $x\in \{x_1,\dots,x_n\}$ is defined as
$$MEI_{\{x_1,\dots,x_n\}} (x) =\dfrac{1}{n} \sum_{i=1}^n \dfrac{\lambda\left(\left\{t\in {\cal I} \,\middle|\, x_i(t) \geq x(t) \right\}\right)}{\lambda(\I)}$$
and it stands for the mean proportion of time that $x$ lies below the curves of the sample. As in the case of the MBD, the MEI is a generalization of the epigraph index that accounts for the proportion of curves that lie entirely above $x$ \citep{HalfRegion11}.

It is clear that MBD and MEI are closely related quantities and that investigating the relation between these two measures might shed light on shape characterization of curves. Indeed, consider a curve with a modified epigraph index close to 0.5. That would mean that the curve is located in the center of the sample in terms of level variation. Then, if the curve's shape is similar to the shapes of the rest of the curves in the sample, one may expect to get a high value for its MBD, since the curve should be contained in many bands defined by other curves. However, if one gets a low MBD value, that would indicate that the curve is contained in a small number of bands, even if it is placed in the center of the sample in terms of level. That can only means that the curve exhibits a shape very different from those of the rest of the curves.

In this section we want to investigate the relationship between these two measures in order to characterize shape outlyingness. The following equality gives an expression of MBD in terms of MEI.

\begin{figure}[!p]
\centering
\includegraphics[width=15.5cm]{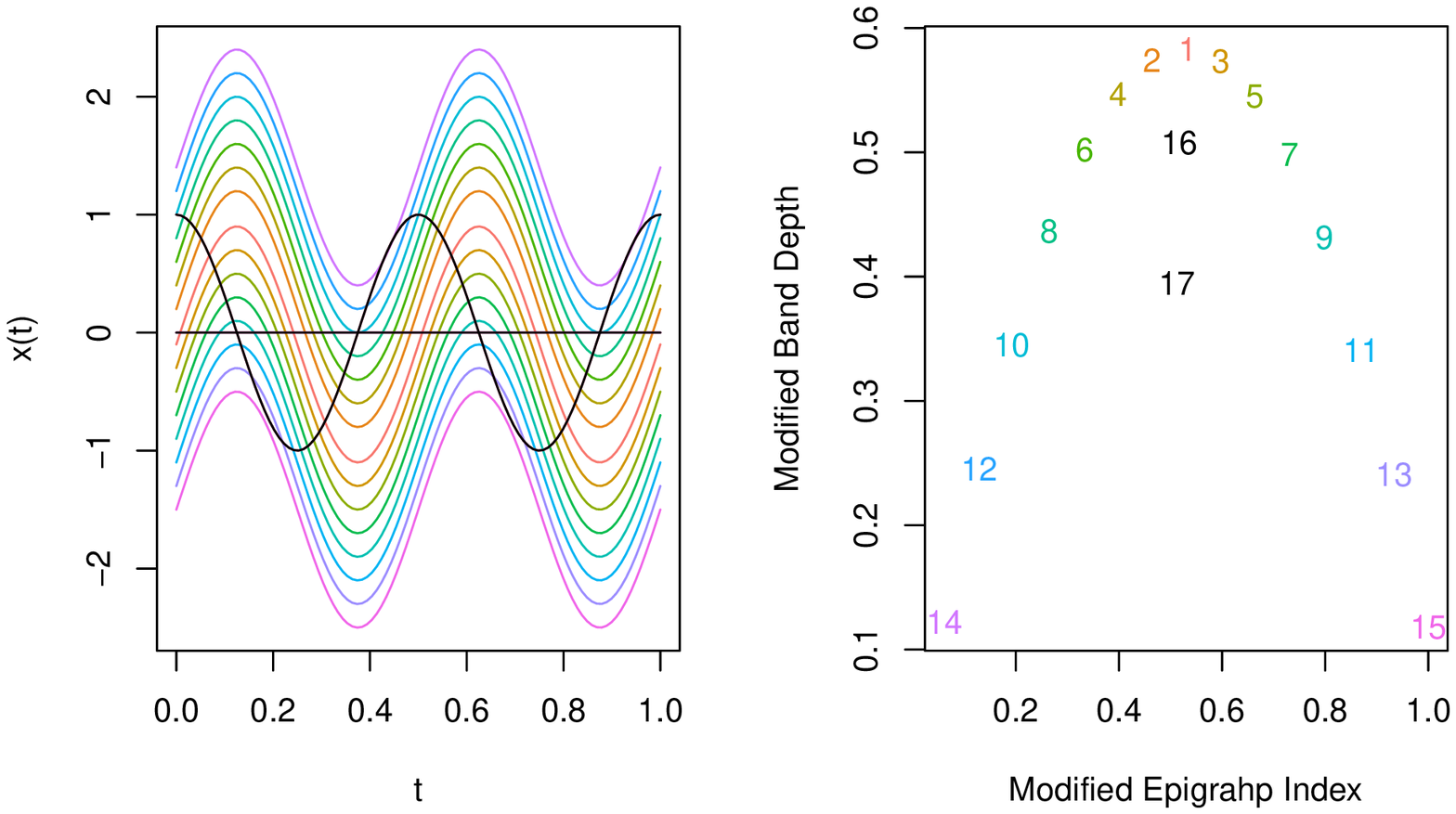}\vspace{-0.5cm}
\caption{Left: A sample of 17 curves of the form $x_i(t)=sin(4\pi t)+(-1)^i\dfrac{i}{10}$, $i=1,\ldots,15$, $x_{16}(t)=0$ and $x_{17}(t)=cos(4\pi t)$, $t\in [0,1]$. Curves $x_{16}$ and $x_{17}$ are displayed in black. Right: Modified band depth versus modified epigraph index of the 17 curves.}
\label{Fex1}
\end{figure}

\begin{figure}[!p]
\centering
\includegraphics[width=15.5cm]{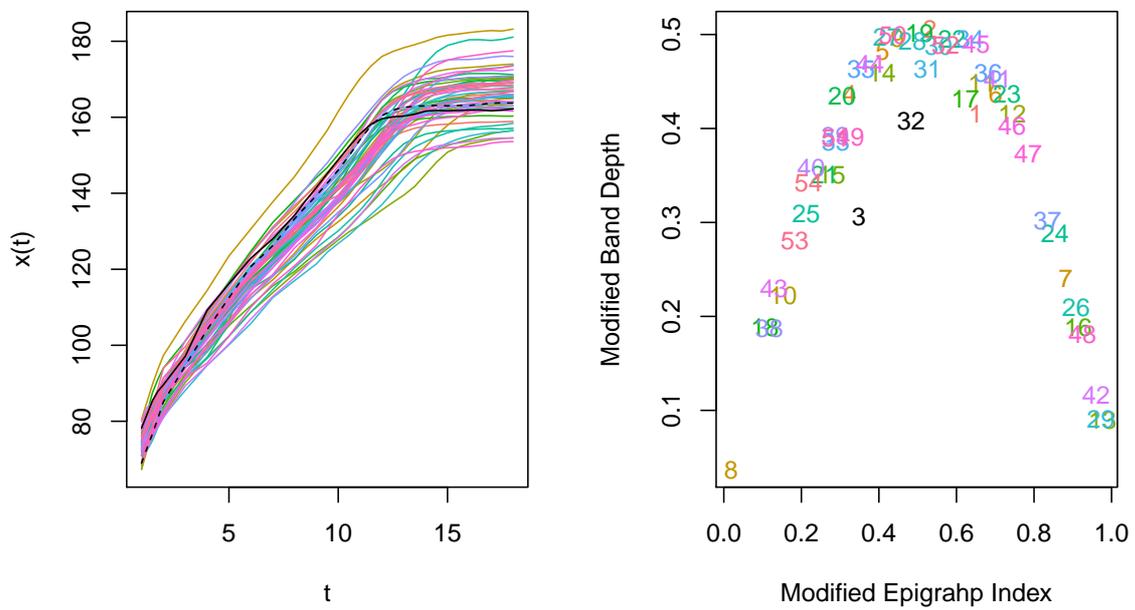}\vspace{-0.5cm}
\caption{Left: Height curves of 54 girls during ages between 0 and 18 years. Curves $x_{3}$ and $x_{32}$ are displayed in black with solid and dashed lines respectively. Right: Modified band depth versus modified epigraph index of the 54 curves.}
\label{Fex2}
\end{figure}

\begin{Proposition}\label{p1} Let $x_1,\dots, x_n$ be $n$ continuous functions on $\I$. Then, for any $x\in \{x_1,\dots,x_n\}$
$$MBD_{\{x_1,\dots,x_n\}} (x) =a_0+a_1 MEI_{\{x_1,\dots,x_n\}} (x) +a_2 \left[ \sum_{i=1}^n \sum_{j=1}^n \dfrac{\lambda(E_{i,x} \cap E_{j,x})}{\lambda(\I)}    \right]                        $$
with $a_0=a_2=\dfrac{-2}{n(n-1)}$, $a_1=\dfrac{2(n+1)}{n-1}$ and $E_{i,x}=\left\{t\in {\cal I} \,\middle|\, x_i(t)\geq x(t)\right\}$.
\end{Proposition}
\begin{proof}
Let $B_{i,j,x}=\left\{t\in {\cal I} \,\middle|\, \min( x_i(t),x_j(t)) \leq x(t) \leq \max( x_i(t),x_j(t))\right\}$. Then we can write\vspace{-0.1cm}
$$MBD_{\{x_1,\dots,x_n\}} (x) =\binom{n}{2}^{-1} \sum_{i=1}^n \sum_{j=i+1}^n \dfrac{\lambda(B_{i,j,x})}{\lambda(\I)}=\dfrac{2}{n(n-1)\lambda(\I)}\sum_{i=1}^n\sum_{j=i+1}^n\lambda(B_{i,j,x}).$$
It is easy to see that $B_{i,j,x} = (E_{i,x} \cap \overline{E}_{j,x})\cup  (\overline{E}_{i,x} \cap E_{j,x})$ if $x\neq x_i, x\neq x_j$ and $B_{i,j,x} =\I$ otherwise, where $\overline{A}$ denotes the complement of set $A$. Then\vspace{-0.1cm}
$$\lambda(B_{i,j,x}) = \left\{ \begin{array}{lll}  \lambda(E_{i,x})+\lambda(E_{j,x})-2\lambda(E_{i,x}\cap E_{j,x}) & \mbox{if}& x\neq x_i, x\neq x_j \\
2\lambda(E_{i,x})+\lambda(E_{j,x})-2\lambda(E_{i,x}\cap E_{j,x}) & \mbox{if}&x=x_i\\
\lambda(E_{i,x})+2\lambda(E_{j,x})-2\lambda(E_{i,x}\cap E_{j,x}) & \mbox{if}&x= x_j\end{array}\right. \vspace{-0.1cm}$$
so we get\vspace{-0.1cm}
\begin{multline*}
\sum_{i=1}^n \sum_{j=i+1}^n \lambda(B_{i,j,x})=\sum_{i=1}^n \sum_{j=i+1}^n[\lambda(E_{i,x})+\lambda(E_{j,x})-2\lambda(E_{i,x}\cap E_{j,x}) ] +\sum_{i=1}^n \lambda(E_{i,x}) -\lambda(\I)\\
=(n-1) \sum_{i=1}^n \lambda(E_{i,x}) -2\sum_{i=1}^n \sum_{j=i+1}^n \lambda(E_{i,x}\cap E_{j,x})  +\sum_{i=1}^n \lambda(E_{i,x}) -\lambda(\I)\\
=(n+1) \sum_{i=1}^n \lambda(E_{i,x}) -2\sum_{i=1}^n \sum_{j=i+1}^n \lambda(E_{i,x}\cap E_{j,x})  -\sum_{i=1}^n \lambda(E_{i,x}) -\lambda(\I)\\
=n(n+1) \dfrac{1}{n}\sum_{i=1}^n \lambda(E_{i,x}) - \sum_{i=1}^n \sum_{j=1}^n \lambda(E_{i,x} \cap E_{j,x})-\lambda(\I) .\vspace{-0.1cm}
\end{multline*}
Then,\vspace{-0.1cm}
\begin{multline*}MBD_{\{x_1,\dots,x_n\}} (x) =\dfrac{2}{n(n-1)\lambda(\I)}\sum_{i=1}^n\sum_{j=i+1}^n\lambda(B_{i,j,x})\\
=\dfrac{-2}{n(n-1)}+\dfrac{2(n+1)}{(n-1)} MEI_{\{x_1,\dots,x_n\}} (x) -\dfrac{2}{n(n-1)} \left[ \sum_{i=1}^n \sum_{j=1}^n \dfrac{\lambda(E_{i,x} \cap E_{j,x})}{\lambda(\I)}    \right] .\vspace{-0.1cm} \end{multline*}
\end{proof}
\begin{Corollary}\label{r1} From Proposition \ref{p1} we get
\begin{multline*}
MBD_{\{x_1,\dots,x_n\}} (x) =a_0+a_1 MEI_{\{x_1,\dots,x_n\}} (x) +a_2 n^2 MEI_{\{x_1,\dots,x_n\}} (x)^2\\+ a_2\left[ \sum_{i=1}^n \sum_{j=1}^n \left(\dfrac{\lambda(E_{i,x} \cap E_{j,x})}{\lambda(\I)}   -\dfrac{\lambda(E_{i,x}) \lambda(E_{j,x})}{\lambda(\I)^2} \right)\right]. \end{multline*}
\end{Corollary}
\begin{Remark}\label{r2} For any sample $x_1,\dots, x_n$ of continuous functions on $\I$, it holds that for any $x\in \{x_1,\dots,x_n\}$,
\begin{equation}\label{parabola}
MBD_{\{x_1,\dots,x_n\}} (x) \leq a_0+a_1 MEI_{\{x_1,\dots,x_n\}} (x) +a_2 n^2 MEI_{\{x_1,\dots,x_n\}} (x)^2. \end{equation}
Indeed,
$$\sum_{i=1}^n \sum_{j=1}^n \left(\dfrac{\lambda(E_{i,x} \cap E_{j,x})}{\lambda(\I)}   -\dfrac{\lambda(E_{i,x}) \lambda(E_{j,x})}{\lambda(\I)^2} \right) = \int_{\I} a_{x}(t)^2 \dfrac{1}{\lambda(\I)}dt - \left(\int_{\I}a_{x}(t) \dfrac{1}{\lambda(\I)}dt \right)^2\geq 0$$
from Jensen's inequality, where $a_x(t)$ is the number of curves above or equal to $x$ at time $t$. Then, since $a_2<0$, (\ref{parabola}) holds from Corollary \ref{r1}.\\
Morover, if none of the curves in the sample cross each other on $\I$, that is, if $(x_i(t_1)-x_j(t_1))(x_i(t_2)-x_j(t_2))>0$ $\forall t_1,t_2 \in \I$, $\forall i\neq j$, then we have for any $x\in \{x_1,\dots,x_n\}$
$$MBD_{\{x_1,\dots,x_n\}} (x) =a_0+a_1 MEI_{\{x_1,\dots,x_n\}} (x) +a_2 n^2 MEI_{\{x_1,\dots,x_n\}} (x)^2,$$
since in that case $E_{i,x}=\emptyset$ or $E_{i,x}=\I$ for all $i$ and so, $\lambda(E_{i,x} \cap E_{j,x})/\lambda(\I)$ and  $\lambda(E_{i,x}) \lambda(E_{j,x})/\lambda(\I)^2$ are equal for all $i,j$.
Then MBD and MEI computed over the curves of the sample define a perfect parabola.
\end{Remark}
An underlying idea below the statement of Remark \ref{r2} is that if in a sample of perfectly aligned curves with  common shape one introduces a curve with a different pattern, then the $\R^2$ point corresponding to the pair $(MEI,MBD)$ for this new curve will lie far away from the parabola defined by the points corresponding to the rest of the curves. Figure \ref{Fex1} provides an example of this phenomenon, in which it is very easy to detect the outlying observations by looking at the MBD vs MEI representation. In general, however, trajectories of a random process will cross many times even if they all exhibit the same trend pattern. Then, the $(MEI,MBD)$ points will not define a perfect parabola and identifying outlying trajectories will not be straightforward. Let us consider as an illustration the height curves of 54 girls coming form the well known Berkeley growth dataset \citep[see, e. g.,][]{RamSil}. They are presented in Figure \ref{Fex2} together with their MBD vs MEI representation. The points corresponding to the $3$rd and $32$nd girls are far from the rest of the points. These two girls exhibit a different growth pattern than the rest of the girls in the sense that they were ones of the highest of the sample during childhood, especially true for girl number 3, but they stopped growing earlier than the rest of the girls and ended up with a height in the lower quartile at 18 year old. Girl number 32, in addition, was one of the smallest girls at birth and exhibits a very high growing rate during the first years of her life. To what extent we can consider these two trajectories as outliers is something that can be addressed by considering the problem of outlier detection in the MBD vs MEI plane as we do in the next section.

\section{Shape outlier detection algorithm and outliergram}\label{salgo}

\begin{figure}[!p]\vspace{-1.5cm}
\centering
\includegraphics[width=7.75cm]{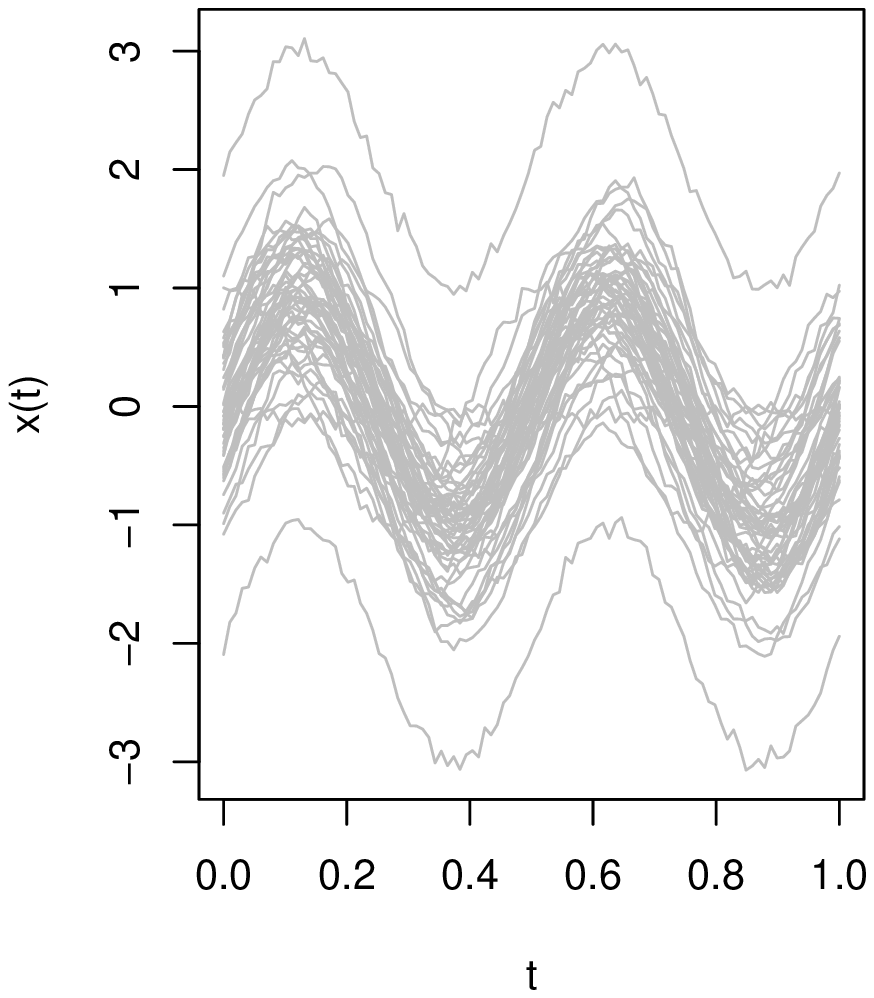}\vspace{-2.5cm}\\
\includegraphics[width=15.5cm]{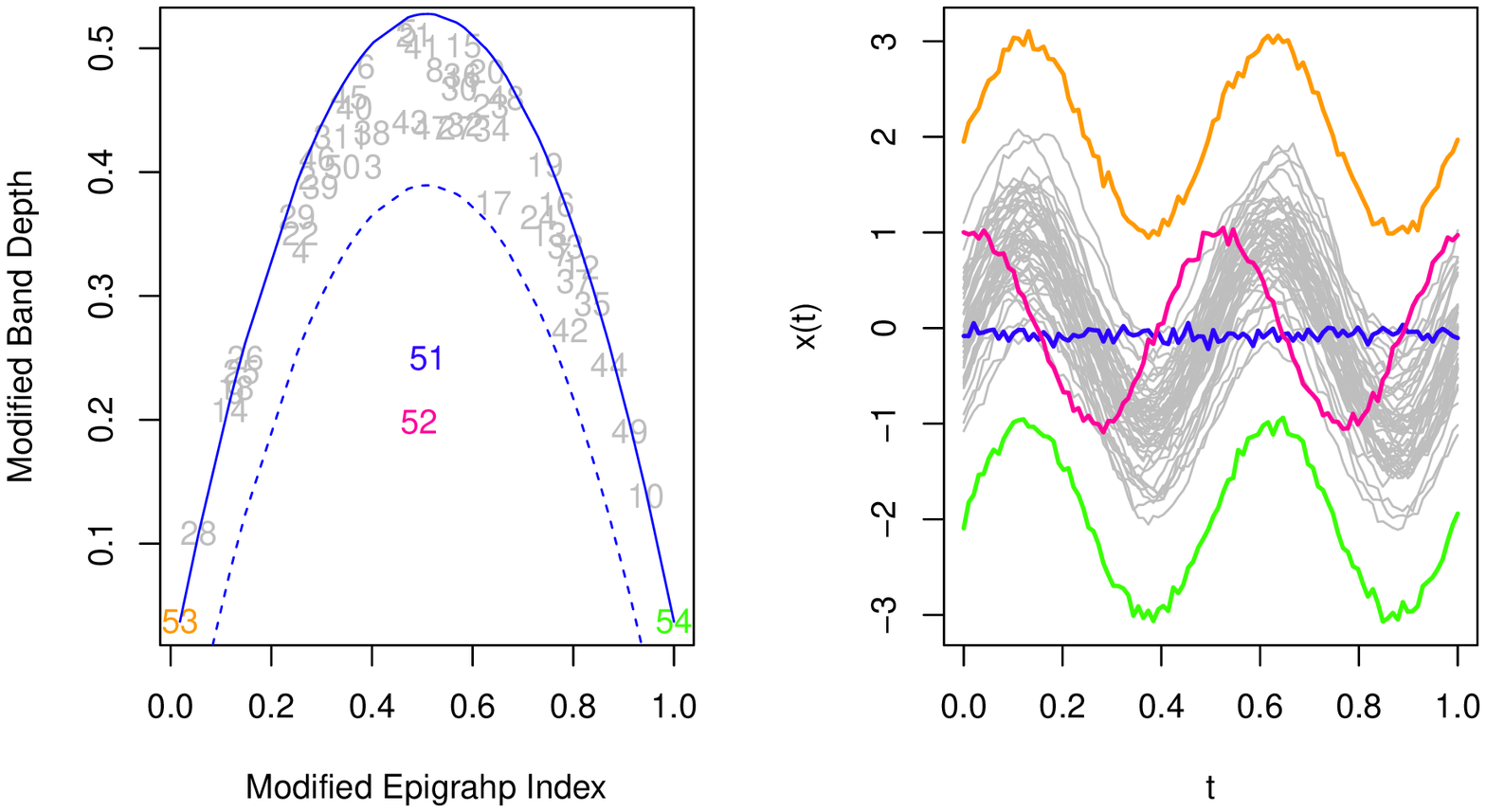}\vspace{-0.5cm}
\caption{Top: $50$ curves generated under the process $X(t)=\sin(4\pi t) +\varepsilon(t)$, $t\in [0,1]$, where $\varepsilon(t)$ is a Gaussian process with zero mean and covariance function $\gamma(s,t)=0.2\exp \{-0.8|s-t|\}$, and four outliers (curves $51$ to $54$). Only two magnitude outliers are visually identifiable. Bottom left: Outliergram. The solid parabola is $P_i=a_0 +a_1 me_i +n^2 a_2 me_i^2$ and the dashed one is $P_i - Q_{d3} -1.5IQR$, which represents the boundary between outlying and non-outlying observations. Bottom right: same as top where magnitude outliers as well as the outliers detected by the outliergram are now shown in color. The color code is the same in both graphics.}
\label{OG}
\end{figure}

Based on the relation between the modified band depth and the modified epigraph index we now propose to use the MBD vs MEI plane as a visualization tool and we give an algorithm to detect shape outliers. From Remark \ref{r2} we know that all the $(MEI,MBD)$ points lie below the parabola given by (\ref{parabola}) and that the closest points to the parabola correspond to curves with typical shape, whereas the most distant ones represent outlying curves in terms of shape. This motivates the use of the univariate boxplot rule for outlier detection on the vertical distances to the parabola. That is, given a sample of curves $x_1, \ldots,x_n$ with $mb_i=MBD_{x_1,\ldots,x_n}(x_i)$ and $me_i=MEI_{x_1,\ldots,x_n}(x_i)$ for $i=1,\ldots,n$, we consider the distances $d_i=a_0 +a_1 me_i +n^2 a_2 me_i^2-mb_i$ and define as shape outliers those curves with $d_i \geq Q_{d3} + 1.5 IQR_d$, where $Q_{d3}$ and $IQR_d$ are the third quartile and inter-quartile range of $d_1,\ldots,d_n$.
In addition to the outlier detection rule, it might be interesting to assess how distant the outliers are from the rest of the sample or to identify curves that might be close to the outlier region although not inside. To jointly visualize the observations in terms of shape and the boundary between the outlying and non-outlying curves we propose to represent in $\R^2$ the $(MEI,MBD)$ points together with the parabola shifted downwards by $Q_{d3} + 1.5 IQR_d$. We will refer to this graphical representation as the outliergram. In Figure \ref{OG} we present an example of such a representation.\\
Although the outliergram has not been conceived to detect magnitude outliers, notice that some particular magnitude outliers, the ones that lie below or above the majority of the curves along most of the time interval, will appear at the bottom corners of the plot. The gap between them and the contiguous observations (according to the order induced by MEI) might be an indicator of their outlyingness. However we do not pretend to give an specific rule to detect them, since very good mechanisms for this purpose already exists, such as the functional boxplot defined in \cite{SunGenton11} \citep[see also][]{HalfBP}. Indeed, we propose to combine that algorithm with our shape outlier detection rule and we provide code that does so (see Supplementary Material information).

\begin{figure}[!p]
\centering
{\footnotesize\hspace{0.6cm}Model 1 \hspace{3.5cm} Model 2  \hspace{3.5cm} Model 3}\\
\includegraphics[width=4.5cm]{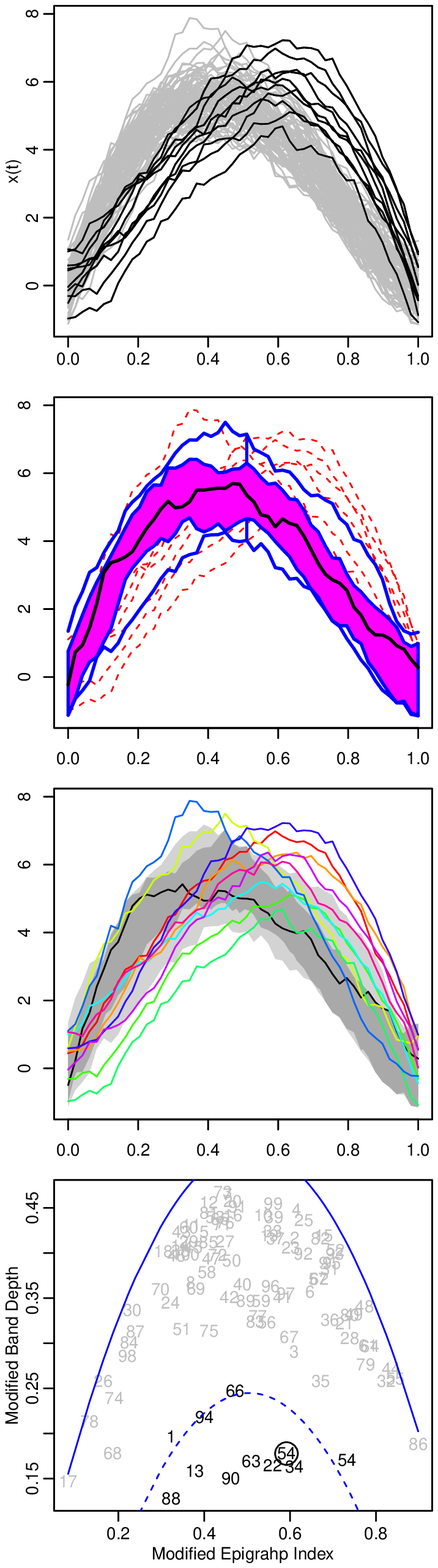}\hspace{-0.1cm}
\includegraphics[width=4.5cm]{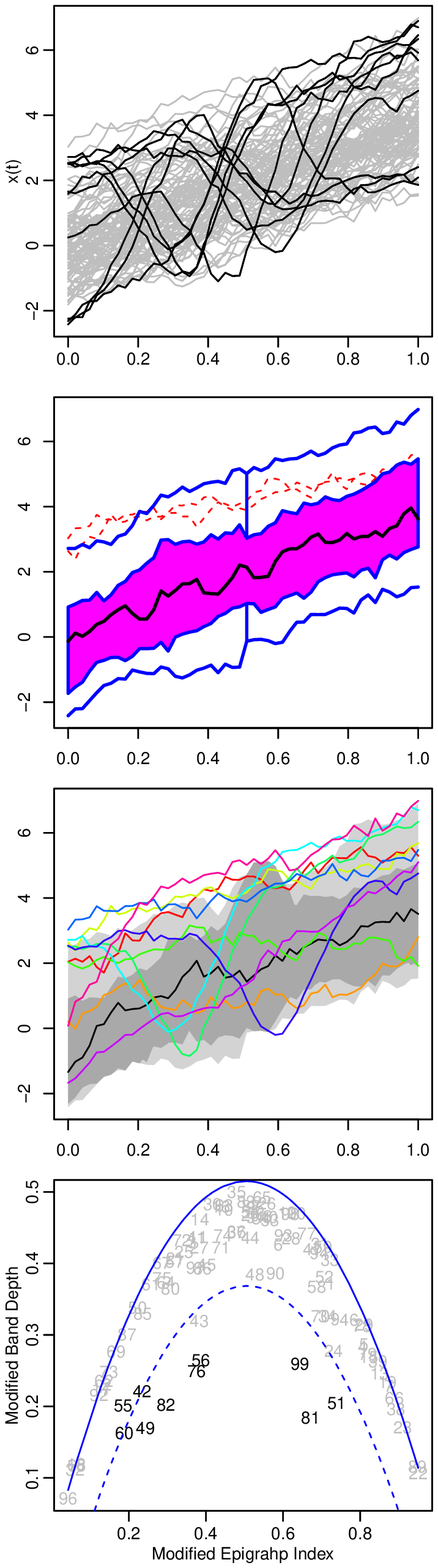}\hspace{-0.15cm}
\includegraphics[width=4.5cm]{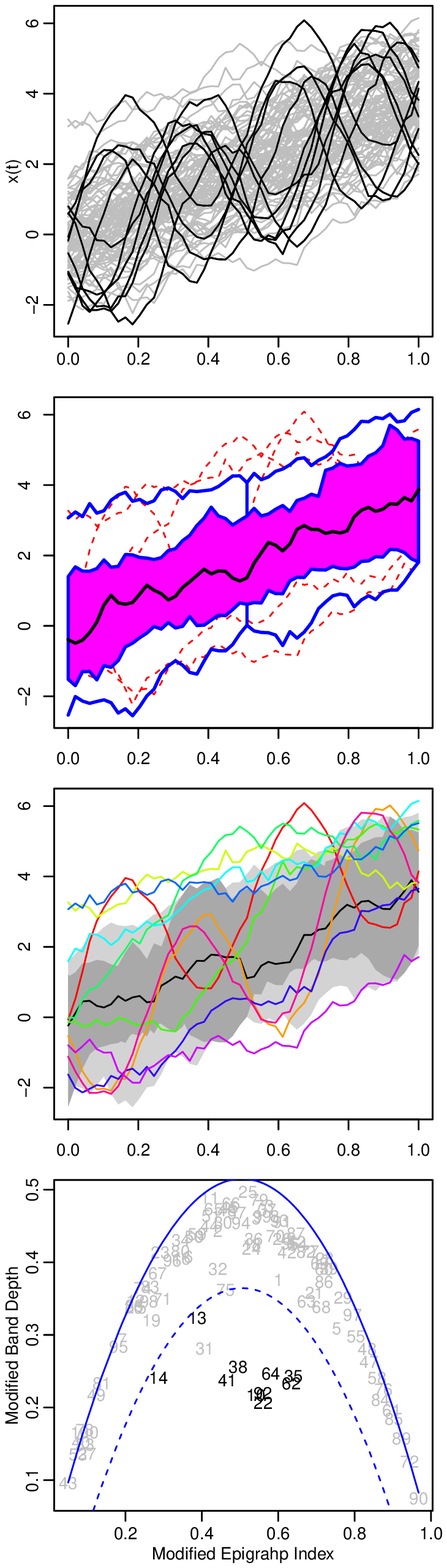}
\caption{Three simulation runs from models 1, 2, and 3 with $n=100$ and $c=0.1$. First row: curves generated from the main model (gray) and the contamination model (black). Second row: Adjusted functional boxplot (outliers are red-dashed lines). Third row: Functional HDR boxplot (outliers are colored lines, the black line is the mode). Fourth row: Adjusted outliergram (outliers correspond to the points below the dashed parabola; the code color is the same of the first row; circles stand for curves that have been considered outliers after having been shifted vertically towards the center of the sample).}
\label{Fsim}
\end{figure}

\begin{figure}[!p]
\centering
{\small\hspace{0.25cm} Girls\hspace{5.5cm} Boys }\\
\includegraphics[width=5cm]{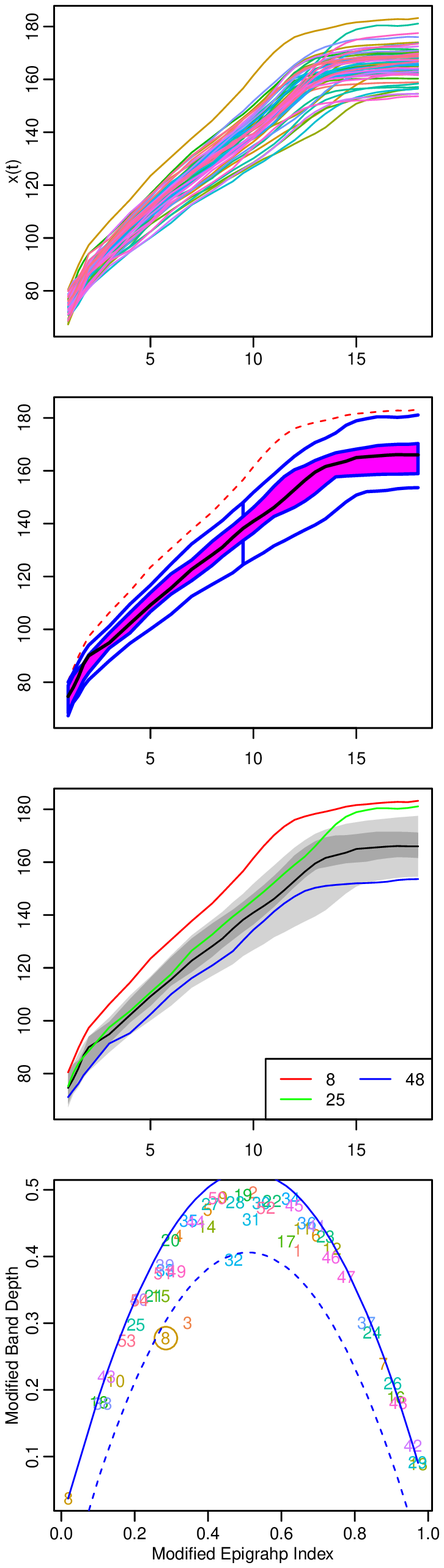}\hspace{1cm}
\includegraphics[width=5cm]{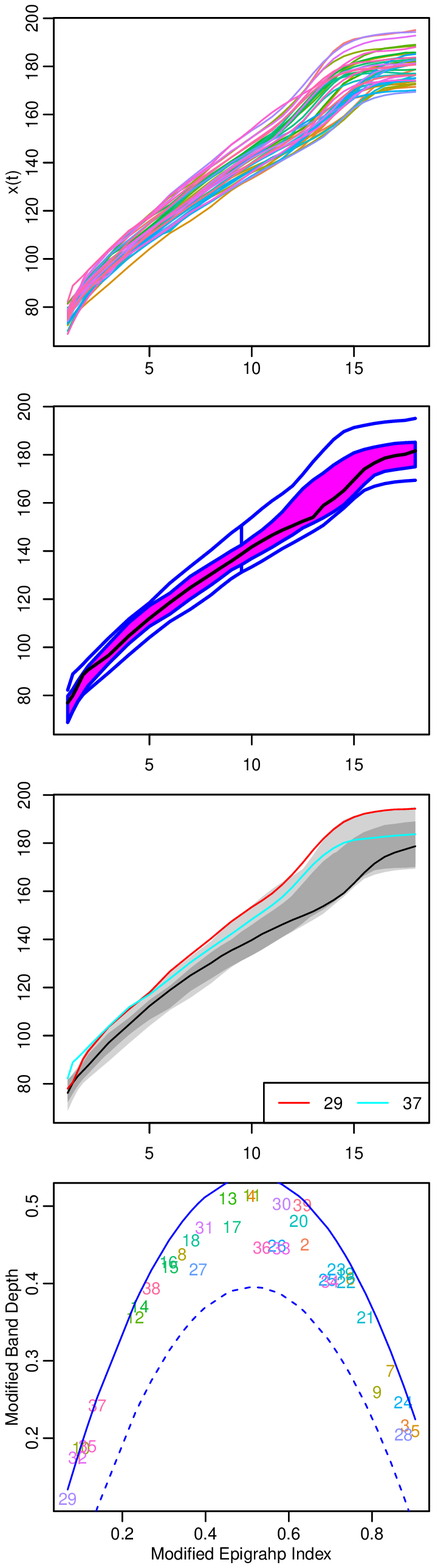}
\caption{Observed curves, adjusted functional boxplot, functional HDR boxplot and adjusted outliergram for the girls (left) and boys (right) growth curves.}
\label{Fgrowth1}
\end{figure}

\begin{figure}[!p]
\centering
{\small\hspace{0.7cm} Raw\hspace{5.05cm} Smoothed }\\
\includegraphics[width=5cm]{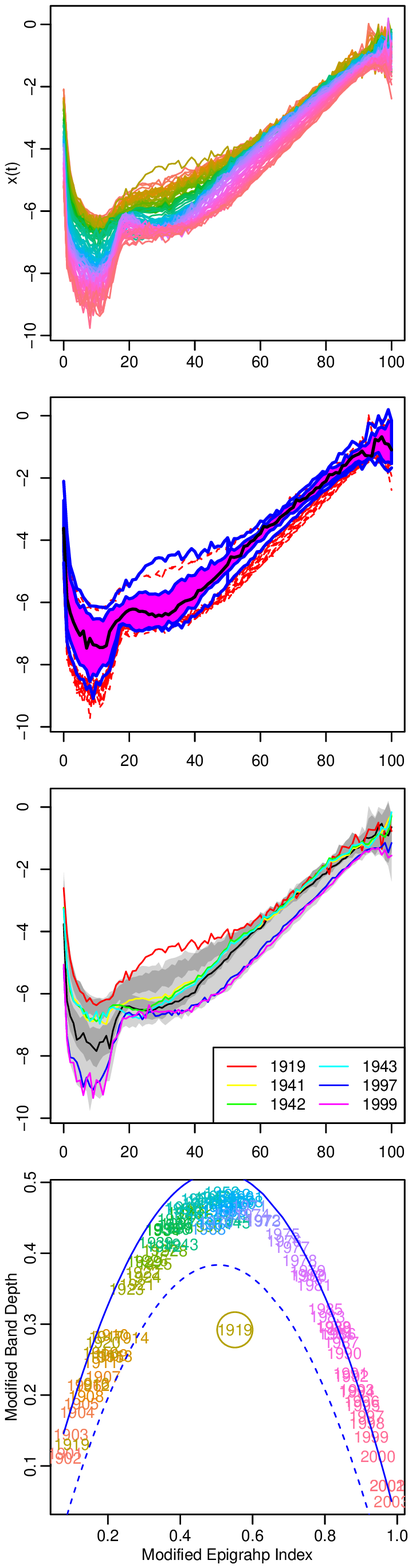}\hspace{1cm}
\includegraphics[width=5cm]{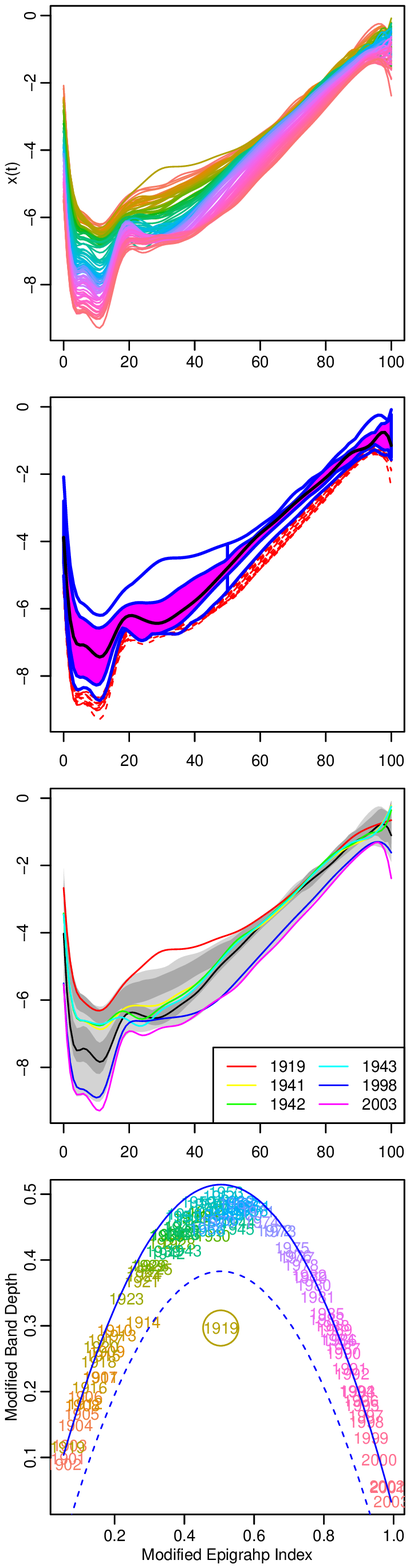}
\caption{Observed curves, adjusted functional boxplot, functional HDR boxplot and adjusted outliergram for raw (left) and smoothed (right) Australian male log-mortality rates.}
\label{FAus}
\end{figure}

It is worth noticing that the precedent reasoning for shape outlier detection might fail with curves that lie above or below the majority of the curves in the sample, that is, with MEI values close to $0$ or $1$. Indeed, for such curves the modified band depth will always be low, since they are surrounded by very few curves, independently of the fact that they might present an atypical shape or not. However, if the curve presented a typical shape and we shifted it vertically towards the center of the sample its MBD in the new location should increase (as MEI increases or decreases). On the other hand, if the curve's shape was atypical, even when placed in the center of the sample, its MBD would remain low. That motivates the addition of a second step in the shape outlier detection procedure in which the more extreme curves (see below for a proper definition) are vertically shifted towards the center of the sample one by one. They would be considered shape outliers if the new (MBD,MEI) point lies in the outlying region previously determined. In that case the outliergram will show both the old and the new (MBD,MEI) points, using a different symbol for the last one to help distinguishing them (see Figures \ref{Fsim}, \ref{Fgrowth1} and \ref{FAus} for some examples).\\
Then, given a sample of curves $x_1, \ldots,x_n$, the whole shape outlier detection algorithm is as follows:

\begin{enumerate}
\item Compute $mb_i=MBD_{x_1,\ldots,x_n}(x_i)$ , $me_i=MEI_{x_1,\ldots,x_n}(x_i)$ and $P_i=a_0 +a_1 me_i +n^2 a_2 me_i^2$, for $i=1,\ldots,n$, where $a_0, a_1$ and $a_2$ are the ones given in Proposition \ref{p1}.
\item Compute $d_i= P_i-mb_i$ for $i=1,\ldots,n$.
\item Compute the third quartile and inter-quartile range of the sample $d_1,\ldots,d_n$, $Q_{d3}$ and $IQR_d$.
\item Shape outlier identification: $SO=\left\{i \,\middle|\, mb_i\leq P_i - Q_{d3} -1.5IQR_d\right\}$.
\item For $i \in \{1,\ldots,n\} \setminus SO$: \begin{itemize}
\item If $x_i(t) < \min_{j\ne i} x_j(t)$ for some $t\in \I$, define $\tilde{x}_i(t)=x_i(t) -\min_t \{x_i(t) - \min_{j\ne i} x_j(t)\}$.
\item If $x_i(t) > \max_{j\ne i} x_j(t)$ for some $t\in \I$, define $\tilde{x}_i(t)=x_i(t) -\max_t \{x_i(t) - \max_{j\ne i} x_j(t)\}$.
\item Compute $\widetilde{mb}_i=MBD_{x_1,\ldots,\tilde{x}_i\ldots,x_n}(\tilde{x}_i)$, $\widetilde{me}_i=MEI_{x_1,\ldots,\tilde{x}_i\ldots,x_n}(\tilde{x}_i)$ and $\widetilde{P}_i=a_0 +a_1 \widetilde{me}_i +n^2 a_2 \widetilde{me}_i^2$. If $\widetilde{mb}_i\leq \widetilde{P}_i - Q_{d3} -1.5IQR_d$ then $SO=SO \cup \{i\}$.
\end{itemize}
\end{enumerate}

\section{Simulation study}\label{ssim}
In this section we compare the performance of the proposed procedure with several functional and/or multivariate outlier detection methods through an extensive simulation study. Namely, we consider eight different techniques developed during the last decade (the numbers below stand only for reference purposes).
\begin{enumerate}
\item Functional boxplot \citep{SunGenton11}: considering the center outward ordering induced by band depth or modified band depth in a sample of curves, a boxplot is constructed by defining the envelope of the $50\%$ central region.The maximum non-outlying envelope is obtained by inflating that central region 1.5 times as in a univariate boxplot. Any curve lying partially or globally outside that maximum non-outlying envelope is considered an outlier.
\item Adjusted functional boxplot \citep{SunGenton12}: The 1.5 constant factor in the functional boxplot (method 1) can be replaced by a different quantity in the aim of controlling the probability of correctly detecting no outliers. The method consists in simulating observations without outliers on the basis of a robust estimator of the covariance function of the data. The factor is then selected as the one for which, when used with the functional boxplot, the probability of detecting no outliers is the closest to 0.993. The selected factor is applied to the functional boxplot of the original data.
\item Functional highest density region (HDR) boxplot \citep{rainbow}:  a functional boxplot is obtained by constructing a bivariate HDR boxplot \citep{Hyndman96} with the first two robust principal component scores. The coverage probability of the outlying region needs to be prespecified.
\item Robust Mahalanobis Distance: considering the curves as multivariate observations, the robust Mahalanobis distance between each curve and the pointwise sample mean is computed. Outliers are defined as observations that have squared robust Mahalanobis distances greater than the $0.99$ quantile of a $\chi^2$ distribution with $p$ degrees of freedom, where $p$ is the fixed number of observation points in every curve \citep[see][and the references therein for details]{rainbow}.
\item Integrated Squared Error \citep{HU}: the integrated squared error between each curve in the sample and its projection into a given number $K$ of robust principal components is computed. Outliers are defined as those observations with an integrated squared error greater than a threshold. Throughout this simulation study $K$ is chosen to be equal to 2 and the threshold is set to $s+3.29\sqrt{s}$, where $s$ is the median of the observed ISEs, as suggested in \cite{rainbow}.
\item[6, 7.] Depth based weighting and trimming \citep{FebGalMant08}: considering different depth measures for functional data the authors proposed to define as outliers the curves whose depth levels are below a cutoff. The cutoff is determined by a bootstrap procedure based either on trimming or weighting of the sample. In the first case, the proportion of potential outliers, which is used as trimming level, needs to be prespecified.
\item[8.] Projection based trimming \citep{FraimanSvarc13}: the authors propose a multivariate and functional robust estimation procedure that provides an outlier detection method as a by-product. The method consists on trimming the sample based on random projections. The maximum proportion of observations to be trimmed has to be prespecified.
\end{enumerate}
As for the outliergram presented in this article, we are going to consider two versions: the one described in Section \ref{salgo} (referred to as 9) and an adjusted version (referred to as 10) inspired by \cite{SunGenton12} (see method 2). The idea of this adjusted algorithm is to select the factor that determines the boundary for outlying points in terms of the observed data, in order to be able to control the false positive rate. As in \cite{SunGenton12} we propose to robustly estimate the covariance matrix of the data and simulate data without outliers on the basis of a centered Gaussian process with estimated covariance function. By simulating a large enough number of data sets and applying to each one the outliergram with different factor values, we can then select the factor whose false detection rate is the closest to $0.007$. Though this procedure could be applied to determine the value of $F$ in $Q_{3d}+F\times IQR_d$ we prefer to use as limiting rule $d_i\geq F\times Q_{1d}$ and choose the value of $F$ as described above. The reason for this choice is the following: the distribution of points $d_i$ is right-skewed and outliers only appear at the right tail of it. Then, assuming the observed data contain outliers, $Q_{d3}$ and $IQR_d$ in the uncontaminated simulated samples would probably be smaller than their counterparts in the original sample, whereas it is reasonable to think that $Q_{1d}$ won't significantly change. Hence, it is advisable to use the rule $d_i\geq F\times Q_{1d}$ for the selection of the factor on the basis of uncontaminated samples and for its posterior application to the outliergram of the observed data. In both cases the set of candidate values for the factor needs to be specified. While for the boundary $Q_{3d}+F\times IQR_d$ it seems logical to search for $F$ around $1.5$, it is not clear what the interval of candidate values should be when using the boundary $F\times Q_{1d}$. Given the distribution of points $d_i$ of the data set to be analyzed, it is reasonable to think that in the uncontaminated data sets generated from the estimation of its covariance structure the limit between outlying and non outlying observations could be somewhere between $Q_{3d}$ and $c\max d_i$, where $c>1$ to account for the case in which the original data set contains no outliers itself. Then, in the adjusted outliergram we will let the interval of candidate values for $F$ vary for each data set to be analysed: after computing the quantities $d_i$, the interval will be set to $[Q_{3d}/Q_{1d},c\max d_i /Q_{1d}]$. See below for the particular values used through the simulation study.\\

For the sake of clarity and conciseness we restrict our simulation study to these ten methods. Comparison between these and some other related nonparametric procedures can be found in \cite{rainbow}.\\

We have generated curves from three different models which are described next. In each case, $n-\lceil c\cdot n \rceil$  curves were generated according to the main model and the remaining $\lceil c\cdot n \rceil$ curves according to the contamination model, where for a real number $x$, $\lceil x \rceil$ is the smallest integer not less than $x$.
\begin{itemize}
\item Model 1. Main model: $X(t)= 30 t (1-t)^{3/2} + \varepsilon(t)$, and contamination model: $X(t)= 30 t^{3/2} (1-t) + \varepsilon(t)$, where $t\in [0,1]$ and $\varepsilon(t)$ is a Gaussian process with zero mean and covariance function $\gamma(s,t)=0.3 \exp \{-|s-t|/0.3\}$. This model had already been used in \cite{FebGalMant08} and \cite{FraimanSvarc13}.
\item Model 2. Main model: $X(t)= 4t + \varepsilon(t)$, and contamination model: $X(t)= 4t+(-1)^u 1.8+\dfrac{1}{\sqrt{2\pi 0.01}}\exp\{-(t-\mu)^2/0.02\}+ \varepsilon(t)$, where $t\in [0,1]$, $\varepsilon(t)$ is a Gaussian process with zero mean and covariance function $\gamma(s,t)=\exp \{-|s-t|\}$, $u$ follows a Bernoulli distribution with probability $1/2$ and $\mu$ is uniformly distributed in $[0.25, 0.75]$. The main model had already been used in \cite{SunGenton11}.
\item Model 3. Main model: $X(t)= 4t + \varepsilon(t)$, and contamination model: $X(t)= 4t+2\sin(4(t+\theta)\pi)+ \varepsilon(t)$, where $t\in [0,1]$, $\varepsilon(t)$ is a Gaussian process with zero mean and covariance function $\gamma(s,t)=\exp \{-|s-t|\}$ and $\theta$ is uniformly distributed in $[0.25, 0.75]$
\end{itemize}
For each one of the three models we considered two different values for the sample size, $n=100$ and $200$, and five values for the contamination rate, $c= 0, 0.05, 0.1, 0.15$ and $0.2$, including the non contaminated model ($c=0$). We ran 400 simulations for each combination of $n$ and $c$. In each case we sampled the curves on $50$ equidistant points over the interval $[0,1]$. For the procedures requiring the specification of the coverage probability of the outlying region or trimming proportion we set those equal to the true value $c$. For procedures $6$ and $7$ we set to $200$ the number of bootstrap samples and chose as depth function the modal depth \citep{ModalDepth} as advised in \cite{FebGalMant08}. For the procedures 2 and 10 we equally set to 200 the number of sampled data sets for the selection of the factor in the outlier detection rule. For these methods we used as robust estimator of the covariance the orthogonalized Gnanadesikan-Kettenring estimator proposed by \cite{MaronnaZamar02}. For method 2, the search interval for the factor value was $[0.5,2.5]$, with a distance between candidate values of 0.25. For method 10, the search was performed in a grid of 10 points equally spaced on the interval $[Q_{d3}/Q_{d1},1.5\max_i d_i/Q_{d1}]$, where $d_1,\ldots,d_n$ refer to the distances on the original data set. For procedure $8$ we set to $100$ the maximum number of random projections. In Figure \ref{Fsim} we present the curves generated with each of these models in a single simulation run.
In Tables \ref{Tsim1} and \ref{Tsim2} we show the results for the two versions of the outliergram and the other eigth methods in terms of the proportion of correctly identified outliers $p_c$ (number of correctly identified outliers over the number of outliers in the sample) and the proportion of false positives $p_f$ (number of wrongly identified outliers over the number of non-outlying curves in the sample). For model 1, the method that achieves the best performance is the Robust Mahalanobis Distance whose $p_c$ and $p_f$ remain very high and low respectively across different sample sizes and contamination rates. However, this method's sensitivity for models 2 and 3 is very low. On the contrary, the ISE method performs very well on models 2 and 3 and presents slightly worse results for model 1. Let us point out that, although the sensitivity of this method is in general very high, its false detection rate is often larger than that of most of the methods. The functional boxplot has very low sensitivity in all the models, although its adjusted version correctly identifies a considerably larger proportion of outliers while barely increasing its false detection rate. The HDR functional boxplot detects more outliers than both versions of the functional boxplot, for model 1, and slightly less than the adjusted version for models 2 and 3, but in general it exhibits very large false detection rates. With respect to the depth based methods (6 and 7), the trimming procedure works always better than the weighting procedure (except for $c=0.05$, where their performances are similar) and they both present better results for $n=200$ than for $n=100$. Except for model 1, they seem to be quite resistant since their sensitivity remains almost constant as the contamination rate increases. That is also the case for the projection based procedure, whose sensitivity is, however, generally low and its false detection rate quite large across all models, sample sizes and contamination rates. With respect to the outliergram we can see that it presents very high detection rates in the three models, especially for $c=0.05$ and $c=0.1$. For models 2 and 3 the sensitivity remains high for larger contamination rates, but for model 1 it decreases rapidly as the contamination rate increases (especially when going from 0.15 to 0.2). Indeed, the outliergram is not a resistant procedure, since the presence of too many outlying trajectories would make decrease the MBD of all the curves in the sample making the (MEI,MBD) representation too spread to find outliers. With regard to specificity, the outliergram presents high false detection rates particularly in the presence of none or few outliers, whereas they decrease as contamination rates increase. The adjusted outliergram however reduces significantly the proportion of incorrectly identified outliers with values close to the nominal level, 0.007, in the case of uncontaminated samples. This comes at the price of a slightly reduced sensitivity, especially in model 1. Finally, let us recall that the HDR functional boxplot, the depth based trimming method and the projection based method had the advantage of having been given the \emph{true} proportion of outliers in each sample.\\
The simulations have been conducted in R using the functions implemented in the packages \texttt{fda.usc} \citep[methods 5 and 6]{fdausc} and \texttt{rainbow} \citep[methods 2, 3 and 4]{rainbowR}. The R-code for a fast version of the functional boxplot \citep{SunGenton11} is available at the second author website \href{http://www.stat.tamu.edu/~sunwards/publication.html}{http://www.stat.tamu.edu/~sunwards/publication.html}, although the general function can also be found in the \texttt{fda} package \citep{fda}. We have obtained the code for the projection based trimming method \citep{FraimanSvarc13} from the authors. The R-code for our shape outlier detection method (standard and adjusted versions) can be found in the supplementary materials. Its implementation relies on a fast computation of the modified band depth and modified epigraph index as in the R package \texttt{Depth.Tools} (\citealp{DepthTools};\citealp[ see also][]{FastBD}). The orthogonalized Gnanadesikan-Kettenring estimator used in methods 2 and 10 is implemented in the R package \texttt{robustbase} \citep{robustbase}. In Table \ref{time} we show the computing time required by each of the methods to run on a sample of $200$ curves in the implementations mentioned before and with the parameter settings used through the simulation study  (R 3.0.1 on a Mac OS X 10.8.4, 2.5 GHz, 4GB of RAM). For the functions that produce graphics, the running time with the graphic option disabled is considered. Notice that all the methods are running under implementations publicly available or provided by their authors except for the adjusted functional boxplot, which we have implemented in a suboptimal way relying on succesive callings to the available function for fast functional boxplots.

\section{Applications}\label{sapp}

In this section we conduct outlier detection in two real data sets with the aim of analysing qualitative differences between the outliergram and previous existing methods and illustrating the use and interpretation of the outliergram in practice.

\subsection{Berkeley growth data}
The well known Berkeley growth data set contains height curves from 54 girls and 39 boys measured at 31 fixed time points between 0 and 18 year old \citep[see][for details]{RamSil}. This dataset is publicly available in R in the \texttt{fda} package \citep{fda}, among others. An outlier detection analysis performed in each group is summarized in Table \ref{growth1}, where we present the different outliers detected in each group by the different methods described in the previous section. For the methods requiring the specification of an expected outlier rate the value 0.05 has been considered. In Figure \ref{Fgrowth1} we show the original curves and graphical representations for three of the methods: the adjusted functional boxplot, the functional HDR boxplot and the adjusted outliergram. Notice that the outliergram and adjusted outliergram representations only differ in the position of the boundary (dashed parabola) between outlying and non outlying observations, the $(MEI,MBD)$ points beeing the same in both cases. In the girls sample, curve $8$, the curve corresponding to the tallest girl, is identified by all the methods. While this curve may be considered a magnitude outlier, notice that it is also detected by the outliergram at step 5 of the algorithm described in Section \ref{salgo}. That is, when shifted towards the center of the sample, the curve is considered a shape outlier since its slopes in the different growth phases are always higher than those of the rest of girls. The other two curves detected by the outliergram are number $3$ and $32$ (see Figure \ref{Fex2}) on which we have already commented in Section \ref{sdepth}.\\
For the boys sample there are five methods that do not detect any outlier and there is no consensus for the rest of the methods. The outliergram detects as outliers curves $9$ and $28$, but the adjusted outliergram does not consider them as atypical observations. Indeed, if we look at the adjusted outliergram of the boys sample we can see how points $9$ and $28$ are the most distant from the solid parabola. However, they are not significantly distant of the rest of the points (as opposed to what happens in the girls sample) to be considered outliers, so when one adjusts the limiting rule for the outliergram, they lie inside the non-outlying region. It is important to notice that in this case visual inspection of the $(MEI,MBD)$ plane may provide more information than a particular limiting rule.\\
As a final comment, is seems clear that in this data set the Integrated Squared Error method detects too many outliers in both samples.

\subsection{Male mortality rates in Australia 1901-2003}
This data set contains the log-mortality rates for the Australian male population between 1901 and 2003. Each log-mortality curve is defined for ages ranging from 0 to 100 years. This data set comes from The Australian Demographic Data Bank and is publicly available in the R package \texttt{fds} \citep{fds}. Since the raw data are very irregular, we have smoothed the curves and conducted the outlier detection analysis in both raw and smoothed data. The results are summarized in Table \ref{aus}. In Figure \ref{FAus} we present the original and smoothed curves and graphical representations for three of the methods in each case. Most of the methods detect as outliers the curves corresponding to the last years of the sample, the 1990's and early 2000's. While it is true that for these years the mortality rates were the lowest of the whole sample, it is difficult to both think of or visually appreciate a significant change or gap between these curves and the corresponding to the immediately preceding years.\\
According to \cite{Ausmort}, there is an evident increase in mortality in 1919, most noticeable in the 15 to 54 years age range, due to the influenza epidemic. Surprisingly, only 5  out of the 10 methods point out the corresponding log-mortality rate curve as an outlier in the analysis of the raw data, and also 5 out of 10 methods in the case of smoothed curves (with some changes). In particular, the corresponding point clearly stands out in the outliergram representation once the curve has been shifted towards the center of the sample, and thus, it is detected as an atypical observation in both the standard and the adjusted version. The standard outliergram also detects some other years as having an atypical mortality rate curve but these are not confirmed by the adjusted outliergram. Indeed, looking at the graphical representation helps discarding this possibility.\\
As for the differences between the analysis performed in raw and smoothed data, it seems clear that all the methods tend to detect more outliers in the first case. Indeed, noise may introduce artificial variation into the sample. With respect to the outliergram, however, even if the standard version detects three more outliers in the raw data than in the smoothed sample, the $(MEI,MBD)$ representation is quite stable across the two samples, and the output of the adjusted outliergram is the same in both cases.

\section{Discussion}\label{sdisc}

This article proposes the outliergram as a tool for representing functional observations in the plane as function of two depth indices, the modified band depth and the modified epigraph index. It allows to visually assess sample variability in terms of shape and to detect potential shape outliers.

The more similar and smooth the curves in the sample, the closer to the parabola (\ref{parabola}) the points in the outliergram. On the other hand, the more noisy the curves and the larger number of crossing points between them, the more dispersed the points under (\ref{parabola}) in the outliergram. In both cases, the points with the largest distances to the parabola represent the most outlying curves, in terms of shape, of the sample. However, it will be easier to detect them if the rest of the points in outliergram are concentrated near (\ref{parabola}). Thus, we suggest applying an smoothing step to noisy data sets before using the outliergram in order to enhance shape differences and similitudes among curves.

In addition to the visualization tool we propose a general boxplot-based rule on the distances to the parabola (\ref{parabola}) to classify observations into outlying and non-outlying in terms of shape. While this is a simple rule with no extra computational cost once the distances to the parabola are obtained, it may tend to detect more outliers than there actually are in the sample. A more sophisticated data set dependent rule is also proposed, where the boundary for the outlying region is chosen to control the false detection rate. According to the simulation results this rule exhibits better specificity and equivalent sensitivity rates than the simple one. However, its computational cost is higher, and sometimes a visual inspection of the outliergram points will be enough to provide good understanding of the nature of the sample and the classifying rule.

For curves that lie above or below the majority of the curves it will be difficult to assess whether they have an atypical shape since they are not surrounded by other curves to which they could be compared to. We propose to shift these curves towards the center of the sample to see if in this new position they stand out as having a different shape. Although we give here a specific rule on which and how much curves should be shifted, extensions to different rules that may cope with the particular nature of different data sets are possible.

Finally, we suggest to combine the outliergram with the functional boxplot \citep{SunGenton11} to account for both magnitude and shape outliers, as we do in the R functions that we provide as Supplementary Material.

\section{Supplementary Material}

The following supplementary material is available online at
\href{http://biostatistics.oxfordjournals.org}%
{http://biostatistics.oxfordjournals.org}.
\begin{description}
\item[R-code:] R-functions for the shape outlier detection algorithm and visualization tool described in this paper (OutGram.R) and for its adjusted version (OutGramAdj.R). These functions also incorporate the functional boxplot technique of \cite{SunGenton11} so that shape and magnitude outlier detection can be combined. However, in the results presented for both the simulation study and the analysis of the two real data sets only the shape outliers have been considered. We also include five other R files with an example of use (example.R, which replicates Figure \ref{OG}), the code for replication of the simulation study conducted in Section \ref{ssim} (simus.R), the code for replication of the analysis of growth and mortality data conducted in Section \ref{sapp} (growth\_app.R and mort\_app.R) and the code to evaluate the computing time of the different methods (runningtime.R). All files can be found in a single zip file (R\_CODE\_OutlierGram.zip).
\end{description}

\section*{Acknowledgments}

The authors acknowledge financial support from grant ECO2011-25706, Spain. The first author also acknowledges financial support from grant MTM2010-17323, Spain.
We are grateful to Marcela Svarc who provided us with the R code for the implementation of the method described in \cite{FraimanSvarc13}.
{\it Conflict of Interest}: None declared.

\bibliographystyle{biorefs}
\bibliography{bib_OutlierGram}


\begin{table}
\centering
{\tiny
\begin{tabular}{lcccccc}
\hline
$n=100$, $c=0$ &\multicolumn{2}{c}{Model 1} &\multicolumn{2}{c}{Model 2}&\multicolumn{2}{c}{Model 3}\\
\hline
Method & $p_c$ & $p_f$ &  $p_c$ &$p_f$ & $p_c$ & $p_f$ \\
  \hline
1. Fun. BP           &- & 0.001 (0.003) & - & 0.001 (0.002) & - & 0.001 (0.003) \\
2. Adj. Fun. BP      &- & 0.006  (0.01) & - & 0.005 (0.009) & - & 0.007 (0.011) \\
3. Fun. HDR BP       &- & 0.000     (0) & - & 0.000     (0) & - & 0.000     (0) \\
4. Rob. Mah. Dist.   &- & 0.016 (0.015) & - & 0.015 (0.014) & - & 0.016 (0.015) \\
5. ISE               &- & 0.035 (0.019) & - & 0.033 (0.018) & - & 0.031 (0.018) \\
6. DB trimming       &- & 0.013 (0.008) & - & 0.012 (0.007) & - & 0.013 (0.007) \\
7. DB weighting      &- & 0.014 (0.012) & - & 0.014 (0.012) & - & 0.015 (0.012) \\
8. PB trimming       &- & 0.000     (0) & - & 0.000     (0) & - & 0.000     (0) \\
9. Outliergram       &- & 0.053 (0.024) & - & 0.054 (0.024) & - & 0.054 (0.023) \\
10. Adj. Outliergram &- & 0.011 (0.012) & - & 0.011 (0.012) & - & 0.012 (0.013) \\
\hline
$n=100$, $c=0.05$ &\multicolumn{2}{c}{Model 1} &\multicolumn{2}{c}{Model 2}&\multicolumn{2}{c}{Model 3}\\
\hline
Method & $p_c$ & $p_f$ &  $p_c$ &$p_f$ & $p_c$ & $p_f$ \\
\hline
1. Fun. BP           &0.194 (0.208) & 0.000 (0.002) & 0.193 (0.191) & 0.001 (0.002) & 0.186 (0.188) & 0.000 (0.002) \\
2. Adj. Fun. BP      &0.574   (0.3) & 0.008 (0.011) & 0.604 (0.294) & 0.007  (0.01) & 0.640 (0.284) & 0.008 (0.011) \\
3. Fun. HDR BP       &0.644 (0.216) & 0.019 (0.011) & 0.474 (0.209) & 0.028 (0.011) & 0.198 (0.178) & 0.042 (0.009) \\
4. Rob. Mah. Dist.   &0.973 (0.078) & 0.009  (0.01) & 0.366 (0.239) & 0.010 (0.012) & 0.100 (0.142) & 0.012 (0.013) \\
5. ISE               &0.906 (0.244) & 0.032 (0.018) & 1.000     (0) & 0.034 (0.021) & 1.000     (0) & 0.031 (0.019) \\
6. DB trimming        &0.95 (0.157) & 0.008 (0.009) & 0.995 (0.031) & 0.007 (0.007) & 1.000     (0) & 0.005 (0.007) \\
7. DB weighting      &0.906 (0.212) & 0.007 (0.008) & 0.994 (0.034) & 0.011  (0.01) & 1.000     (0) & 0.012 (0.011) \\
8. PB trimming       &0.375 (0.199) & 0.033  (0.01) & 0.394 (0.193) & 0.032  (0.01) & 0.185 (0.156) & 0.043 (0.008) \\
9. Outliergram       &0.998 (0.017) & 0.035 (0.019) & 1.000     (0) & 0.034  (0.02) & 1.000     (0) & 0.034 (0.021) \\
10. Adj. Outliergram &0.983 (0.063) & 0.008 (0.011) & 0.984 (0.063) & 0.008  (0.01) & 1.000     (0) & 0.009 (0.012) \\
\hline
$n=100$, $c=0.1$ &\multicolumn{2}{c}{Model 1} &\multicolumn{2}{c}{Model 2}&\multicolumn{2}{c}{Model 3}\\
\hline
Method & $p_c$ & $p_f$ &  $p_c$ &$p_f$ & $p_c$ & $p_f$ \\
\hline
1. Fun. BP           &0.133 (0.139) & 0.000 (0.002) & 0.167 (0.142) & 0.000 (0.002) & 0.170 (0.155) & 0.000 (0.002) \\
2. Adj. Fun. BP      &0.551 (0.256) & 0.006  (0.01) & 0.635  (0.24) & 0.008 (0.011) & 0.661 (0.226) & 0.007  (0.01) \\
3. Fun. HDR BP       &0.624  (0.16) & 0.042 (0.018) & 0.586 (0.134) & 0.046 (0.015) & 0.321 (0.154) & 0.075 (0.017) \\
4. Rob. Mah. Dist.    &0.95 (0.103) & 0.004 (0.007) & 0.367  (0.18) & 0.007 (0.009) & 0.118 (0.134) & 0.010 (0.011) \\
5. ISE               &0.872 (0.277) & 0.026 (0.018) & 1.000     (0) & 0.033 (0.022) & 1.000     (0) & 0.030 (0.019) \\
6. DB trimming       &0.742 (0.371) & 0.009  (0.01) & 0.997 (0.017) & 0.010 (0.009) & 1.000     (0) & 0.006 (0.008) \\
7. DB weighting      &0.124 (0.202) & 0.002 (0.004) & 0.962 (0.092) & 0.006 (0.008) & 0.992 (0.057) & 0.006 (0.008) \\
8. PB trimming       &0.455 (0.136) & 0.061 (0.015) & 0.513 (0.122) & 0.054 (0.014) & 0.295 (0.134) & 0.078 (0.015) \\
9. Outliergram       &0.989 (0.038) & 0.023 (0.017) & 0.998 (0.015) & 0.016 (0.014) & 1.000     (0) & 0.021 (0.016) \\
10. Adj. Outliergram &0.923 (0.106) & 0.005 (0.009) & 0.983 (0.049) & 0.006 (0.009) & 1.000     (0) & 0.008 (0.011) \\
\hline
$n=100$, $c=0.15$ &\multicolumn{2}{c}{Model 1} &\multicolumn{2}{c}{Model 2}&\multicolumn{2}{c}{Model 3}\\
\hline
Method & $p_c$ & $p_f$ &  $p_c$ &$p_f$ & $p_c$ & $p_f$ \\
\hline
1. Fun. BP           &0.091 (0.104) & 0.000 (0.001) & 0.135 (0.114) & 0.000 (0.002) & 0.131  (0.11) & 0.000 (0.002) \\
2. Adj. Fun. BP      &0.498 (0.206) & 0.005 (0.008) & 0.623 (0.208) & 0.006  (0.01) & 0.618 (0.192) & 0.006  (0.01) \\
3. Fun. HDR BP       &0.633 (0.127) & 0.065 (0.022) & 0.669 (0.103) & 0.058 (0.018) & 0.432 (0.134) & 0.100 (0.024) \\
4. Rob. Mah. Dist.   &0.935 (0.119) & 0.001 (0.004) & 0.335 (0.168) & 0.003 (0.007) & 0.136  (0.14) & 0.007  (0.01) \\
5. ISE               &0.753 (0.344) & 0.027 (0.018) & 0.999 (0.011) & 0.041 (0.026) & 1.000     (0) & 0.030  (0.02) \\
6. DB trimming       &0.446 (0.391) & 0.010 (0.012) & 0.998  (0.01) & 0.012 (0.011) & 1.000     (0) & 0.009 (0.009) \\
7. DB weighting      &0.021 (0.038) & 0.001 (0.003) & 0.747 (0.276) & 0.002 (0.005) & 0.682 (0.372) & 0.002 (0.005) \\
8. PB trimming       &0.495 (0.117) & 0.089 (0.021) & 0.576 (0.103) & 0.075 (0.018) & 0.353   (0.1) & 0.114 (0.018) \\
9. Outliergram       &0.895 (0.107) & 0.012 (0.013) & 0.988 (0.033) & 0.008  (0.01) & 1.000 (0.003) & 0.008  (0.01) \\
10. Adj. Outliergram  &0.66 (0.198) & 0.003 (0.006) & 0.967 (0.074) & 0.005 (0.008) & 1.000 (0.005) & 0.008 (0.013) \\
\hline
$n=100$, $c=0.2$ &\multicolumn{2}{c}{Model 1} &\multicolumn{2}{c}{Model 2}&\multicolumn{2}{c}{Model 3}\\
\hline
Method & $p_c$ & $p_f$ &  $p_c$ &$p_f$ & $p_c$ & $p_f$ \\
\hline
1. Fun. BP           &0.048 (0.066) & 0.000     (0) & 0.111 (0.105) & 0.000 (0.002) & 0.100 (0.088) & 0.000 (0.001) \\
2. Adj. Fun. BP      &0.356 (0.205) & 0.003 (0.006) & 0.564 (0.187) & 0.005 (0.008) & 0.567 (0.185) & 0.004 (0.008) \\
3. Fun. HDR BP       &0.607 (0.108) & 0.098 (0.027) & 0.721 (0.088) & 0.070 (0.022) & 0.538 (0.125) & 0.115 (0.031) \\
4. Rob. Mah. Dist.   &0.856 (0.173) & 0.000 (0.001) & 0.299 (0.162) & 0.001 (0.004) & 0.143 (0.143) & 0.005 (0.009) \\
5. ISE               &0.536 (0.387) & 0.027 (0.021) & 0.998 (0.014) & 0.043 (0.026) & 1.000     (0) & 0.029 (0.021) \\
6. DB trimming       &0.168 (0.221) & 0.009 (0.012) & 0.997 (0.017) & 0.012  (0.01) & 0.998 (0.021) & 0.009 (0.009) \\
7. DB weighting      &0.013 (0.027) & 0.001 (0.004) & 0.232 (0.232) & 0.000 (0.003) & 0.096 (0.152) & 0.000 (0.002) \\
8. PB trimming       &0.514 (0.096) & 0.121 (0.024) & 0.619 (0.089) & 0.095 (0.022) & 0.401 (0.086) & 0.150 (0.022) \\
9. Outliergram       &0.367 (0.201) & 0.002 (0.005) & 0.919 (0.134) & 0.002 (0.005) & 0.994 (0.023) & 0.001 (0.004) \\
10. Adj. Outliergram &0.251 (0.163) & 0.002 (0.005) & 0.972 (0.058) & 0.004 (0.007) & 1.000 (0.003) & 0.008 (0.011) \\
\hline
\end{tabular}}
\caption{Mean and standard deviation (in parentheses) of the proportion of correctly and falsely identified outliers in the three simulation models over 400 simulation runs for $n=100$.\label{Tsim1}}
\end{table}

\begin{table}
\centering
{\tiny
\begin{tabular}{lcccccc}
\hline
$n=200$, $c=0$ &\multicolumn{2}{c}{Model 1} &\multicolumn{2}{c}{Model 2}&\multicolumn{2}{c}{Model 3}\\
\hline
Method & $p_c$ & $p_f$ &  $p_c$ &$p_f$ & $p_c$ & $p_f$ \\
\hline
1. Fun. BP           &- & 0.000     (0) & - & 0.000     (0) & - & 0.000 (0.001) \\
2. Adj. Fun. BP      &- & 0.004 (0.006) & - & 0.003 (0.004) & - & 0.004 (0.005) \\
3. Fun. HDR BP       &- & 0.000     (0) & - & 0.000     (0) & - & 0.000     (0) \\
4. Rob. Mah. Dist.   &- & 0.013 (0.009) & - & 0.013 (0.009) & - & 0.012 (0.009) \\
5. ISE               &- & 0.029 (0.013) & - & 0.030 (0.013) & - & 0.030 (0.013) \\
6. DB trimming       &- & 0.012 (0.004) & - & 0.012 (0.004) & - & 0.012 (0.005) \\
7. DB weighting      &- & 0.016 (0.009) & - & 0.016 (0.009) & - & 0.016 (0.008) \\
8. PB trimming       &- & 0.000     (0) & - & 0.000     (0) & - & 0.000     (0) \\
9. Outliergram       &- & 0.049 (0.017) & - & 0.050 (0.017) & - & 0.046 (0.017) \\
10. Adj. Outliergram &- & 0.009 (0.007) & - & 0.009 (0.007) & - & 0.009 (0.007) \\
\hline
$n=200$, $c=0.05$ &\multicolumn{2}{c}{Model 1} &\multicolumn{2}{c}{Model 2}&\multicolumn{2}{c}{Model 3}\\
\hline
Method & $p_c$ & $p_f$ &  $p_c$ &$p_f$ & $p_c$ & $p_f$ \\
\hline
1. Fun. BP           &0.059 (0.084) & 0.000 (0.001) & 0.069 (0.086) & 0.000 (0.001) & 0.061 (0.083) & 0.000     (0) \\
2. Adj. Fun. BP      &0.483 (0.197) & 0.003 (0.005) & 0.501  (0.19) & 0.003 (0.005) & 0.500 (0.209) & 0.003 (0.005) \\
3. Fun. HDR BP       &0.621 (0.148) & 0.020 (0.008) & 0.495 (0.142) & 0.027 (0.007) & 0.183 (0.123) & 0.043 (0.006) \\
4. Rob. Mah. Dist.   &0.982 (0.045) & 0.007 (0.007) & 0.383 (0.178) & 0.009 (0.007) & 0.084 (0.099) & 0.011 (0.009) \\
5. ISE               &0.916 (0.235) & 0.028 (0.012) & 1.000     (0) & 0.029 (0.015) & 1.000     (0) & 0.028 (0.014) \\
6. DB trimming       &0.978 (0.056) & 0.010 (0.005) & 0.998  (0.02) & 0.009 (0.005) & 0.999 (0.011) & 0.008 (0.005) \\
7. DB weighting      &0.959 (0.091) & 0.010 (0.007) & 0.994 (0.048) & 0.014 (0.009) & 0.996 (0.054) & 0.015 (0.009) \\
8. PB trimming       &0.413 (0.132) & 0.031 (0.007) & 0.456 (0.134) & 0.029 (0.007) & 0.216 (0.118) & 0.041 (0.006) \\
9. Outliergram       &0.998 (0.015) & 0.033 (0.013) & 1.000 (0.005) & 0.031 (0.013) & 1.000     (0) & 0.031 (0.013) \\
10. Adj. Outliergram &0.972 (0.057) & 0.006 (0.007) & 0.983 (0.047) & 0.006 (0.007) & 1.000 (0.005) & 0.008 (0.007) \\
\hline
$n=200$, $c=0.1$ &\multicolumn{2}{c}{Model 1} &\multicolumn{2}{c}{Model 2}&\multicolumn{2}{c}{Model 3}\\
\hline
Method & $p_c$ & $p_f$ &  $p_c$ &$p_f$ & $p_c$ & $p_f$ \\
\hline
1. Fun. BP            &0.04  (0.05) & 0.000     (0) & 0.054 (0.056) & 0.000     (0) & 0.058 (0.062) & 0.000     (0) \\
2. Adj. Fun. BP      &0.442 (0.215) & 0.004 (0.006) & 0.520 (0.218) & 0.004 (0.007) & 0.565 (0.231) & 0.005 (0.008) \\
3. Fun. HDR BP       &0.592 (0.106) & 0.045 (0.012) & 0.598 (0.101) & 0.045 (0.011) & 0.324 (0.112) & 0.075 (0.012) \\
4. Rob. Mah. Dist.   &0.974 (0.043) & 0.003 (0.005) & 0.361 (0.137) & 0.005 (0.006) & 0.111 (0.091) & 0.008 (0.008) \\
5. ISE               &0.902 (0.242) & 0.025 (0.012) & 1.000 (0.003) & 0.031 (0.017) & 1.000     (0) & 0.026 (0.013) \\
6. DB trimming       &0.874 (0.273) & 0.011 (0.007) & 0.998 (0.014) & 0.013 (0.007) & 1.000 (0.004) & 0.011 (0.006) \\
7. DB weighting      &0.074 (0.101) & 0.001 (0.003) & 0.984 (0.066) & 0.008 (0.007) & 0.994  (0.07) & 0.008 (0.007) \\
8. PB trimming       &0.469 (0.096) & 0.059 (0.011) & 0.542 (0.091) & 0.051  (0.01) & 0.310 (0.086) & 0.077  (0.01) \\
9. Outliergram       &0.986 (0.028) & 0.022 (0.012) & 0.995 (0.017) & 0.017  (0.01) & 1.000 (0.003) & 0.018 (0.011) \\
10. Adj. Outliergram &0.904 (0.084) & 0.004 (0.006) & 0.967 (0.056) & 0.005 (0.007) & 0.999 (0.007) & 0.006 (0.007) \\
\hline
$n=200$, $c=0.15$ &\multicolumn{2}{c}{Model 1} &\multicolumn{2}{c}{Model 2}&\multicolumn{2}{c}{Model 3}\\
\hline
Method & $p_c$ & $p_f$ &  $p_c$ &$p_f$ & $p_c$ & $p_f$ \\
\hline
1. Fun. BP           &0.023 (0.036) & 0.000     (0) & 0.046 (0.045) & 0.000     (0) & 0.039 (0.043) & 0.000     (0) \\
2. Adj. Fun. BP      &0.369 (0.214) & 0.003 (0.005) & 0.656 (0.231) & 0.007 (0.008) & 0.675 (0.227) & 0.007 (0.008) \\
3. Fun. HDR BP       &0.598 (0.081) & 0.071 (0.014) & 0.661 (0.073) & 0.060 (0.013) & 0.434 (0.101) & 0.100 (0.018) \\
4. Rob. Mah. Dist.   &0.948 (0.072) & 0.001 (0.003) & 0.317 (0.117) & 0.003 (0.004) & 0.118  (0.09) & 0.006 (0.007) \\
5. ISE               &0.779  (0.32) & 0.023 (0.013) & 1.000 (0.003) & 0.038  (0.02) & 1.000     (0) & 0.025 (0.013) \\
6. DB trimming       &0.479 (0.393) & 0.010 (0.008) & 0.999 (0.007) & 0.016 (0.008) & 1.000     (0) & 0.013 (0.007) \\
7. DB weighting      &0.019 (0.027) & 0.001 (0.002) & 0.838 (0.198) & 0.002 (0.004) & 0.726 (0.343) & 0.002 (0.004) \\
8. PB trimming       &0.499 (0.078) & 0.088 (0.014) & 0.590 (0.075) & 0.072 (0.013) & 0.378 (0.075) & 0.110 (0.013) \\
9. Outliergram       &0.889 (0.089) & 0.009 (0.007) & 0.988 (0.025) & 0.008 (0.007) & 1.000 (0.004) & 0.007 (0.006) \\
10. Adj. Outliergram &0.637  (0.15) & 0.002 (0.004) & 0.964 (0.053) & 0.005 (0.006) & 0.999 (0.004) & 0.005 (0.007) \\
\hline
$n=200$, $c=0.2$ &\multicolumn{2}{c}{Model 1} &\multicolumn{2}{c}{Model 2}&\multicolumn{2}{c}{Model 3}\\
\hline
Method & $p_c$ & $p_f$ &  $p_c$ &$p_f$ & $p_c$ & $p_f$ \\
\hline
1. Fun. BP           &0.011 (0.023) & 0.000     (0) & 0.028 (0.033) & 0.000     (0) & 0.026 (0.031) & 0.000     (0) \\
2. Adj. Fun. BP      &0.317 (0.218) & 0.003 (0.005) & 0.679 (0.161) & 0.006 (0.007) & 0.710 (0.143) & 0.007 (0.007) \\
3. Fun. HDR BP       &0.588 (0.074) & 0.103 (0.019) & 0.718 (0.062) & 0.070 (0.015) & 0.521 (0.093) & 0.120 (0.023) \\
4. Rob. Mah. Dist.   &0.878 (0.116) & 0.000 (0.001) & 0.286 (0.108) & 0.001 (0.003) & 0.124 (0.092) & 0.004 (0.006) \\
5. ISE               &0.449 (0.367) & 0.023 (0.014) & 1.000 (0.004) & 0.041 (0.022) & 1.000     (0) & 0.024 (0.014) \\
6. DB trimming       &0.186 (0.231) & 0.008 (0.008) & 0.998 (0.009) & 0.017 (0.007) & 0.999  (0.01) & 0.013 (0.006) \\
7. DB weighting      &0.015 (0.021) & 0.001 (0.003) & 0.199 (0.159) & 0.000 (0.002) & 0.061 (0.062) & 0.000 (0.001) \\
8. PB trimming       &0.521 (0.075) & 0.120 (0.019) & 0.638  (0.06) & 0.091 (0.015) & 0.421 (0.064) & 0.145 (0.016) \\
9. Outliergram       &0.341 (0.164) & 0.002 (0.003) & 0.921 (0.084) & 0.001 (0.003) & 0.994 (0.017) & 0.001 (0.003) \\
10. Adj. Outliergram &0.235 (0.134) & 0.001 (0.003) & 0.964 (0.056) & 0.004 (0.006) & 0.999 (0.005) & 0.008 (0.008) \\
\hline
\end{tabular}}
\caption{Mean and standard deviation (in parentheses) of the proportion of correctly and falsely identified outliers in the three simulation models over 400 simulation runs for $n=200$.\label{Tsim2}}
\end{table}

\begin{table}
\centering
{\footnotesize
\begin{tabular}{lc}
Method & $Time (sec.)$  \\
\hline
1. Functional Boxplot                &   0.050\\
2. Adjusted Functional Boxplot                & 29.801  \\
3. Functional HDR Boxplot        &   2.502\\
4. Robust Mahalanobis Distance&    1.277\\
5. Integrated Squared Error        &  2.132\\
6. Depth based trimming             &   397.691\\
7. Depth based weighting          &   402.899\\
8. Projection based trimming      & 0.067\\
9. Outliergram                         &   2.214\\
10. Adjusted Outliergram        &   6.306\\
\hline
\end{tabular}}
\caption{Computing time (in seconds) for the different methods on a sample of $200$ curves generated under Model 1 with $50$ observation points by curve and contamination rate equal to $0.1$.\label{time}}
\end{table}

\begin{table}
\centering
{\footnotesize
\begin{tabular}{lcc}
Method &Girls sample&Boys sample\\
\hline
1. Functional Boxplot            & $8$                    &     - \\
2. Adjusted Functional Boxplot   & $8$                    &     - \\
3. Functional HDR Boxplot        & $8, 25,48$             & $29, 37$\\
4. Robust Mahalanobis Distance   & $8,13$                 & $37$ \\
5. Integrated Squared Error      & $1, 3, 6, 8, 10, 15, 17, 18, 25,$   & $1, 12, 15, 18, 23, 24,$\\
                                 & $26, 29, 32, 37, 38, 42, 49, 53$    & $26, 34, 35, 36, 37, 38$\\
6. Depth based trimming          & $8$                    & -\\
7. Depth based weighting         & $8$                    &  - \\
8. Projection based trimming     & $8, 13, 25$            & $1, 29$ \\
9. Outliergram                   & $3, 8, 32$             & $9, 28$\\
10. Adjusted Outliergram         & $3, 8, 32$             & -\\
\hline
\end{tabular}}
\caption{Outliers detected by the different methods in the girls and boys height curves samples.\label{growth1}}
\end{table}

\begin{table}
\centering
{\footnotesize
\begin{tabular}{lcc}
Method &Raw data&Smooth data\\
\hline
1. Functional Boxplot   & $1901, 1902, 1990, 1991, 1992, 1993,$ & $1991, 1992, 1993, 1994, 1995,$\\
& $1994, 1995, 1996, 1997, 1998, 1999,$             &     $1996, 1997, 1998, 1999, 2000, 2001,$ \\
& $2000, 2001, 2002, 2003$ & $2002, 2003$\\
2. Adjusted Functional Boxplot   & $1901, 1902, 1990, 1991, 1992, 1993, $ & $1991, 1992, 1993, 1994, 1995,$\\
& $1994, 1995, 1996, 1997, 1998, 1999,$                    &     $1996, 1997, 1998, 1999, 2000, 2001,$ \\
& $2000, 2001, 2002, 2003$ & $2002, 2003$ \\
3. Functional HDR Boxplot        & $1919, 1941, 1942, 1943, 1997, 1999$             & $1919, 1941, 1942, 1943, 1998, 2003$\\
4. Robust Mahalanobis Distance   & $1982, 1991, 1993, 1994, 1997, 1998,$& $1988, 1989, 1990, 1991, 1992, 1993,$ \\
& $1999, 2000, 2001, 2002, 2003$                 & $1994, 1995, 1996, 1997, 1998, 1999,$ \\
& & $2000, 2001, 2002, 2003$\\
5. Integrated Squared Error      & $1913, 1914, 1919, 1998, 1999, 2000,$   & $2001, 2002, 2003$\\
& $2001, 2002, 2003$ & \\
6. Depth based trimming          & $1919, 2003$           & $1919$\\
7. Depth based weighting         & -                      &  - \\
8. Projection based trimming     & $1998, 1999, 2000, 2001, 2002, 2003$            & $1919, 1999, 2000, 2001, 2002, 2003$ \\
9. Outliergram                   & $1901, 1907, 1914, 1915, 1919$             & $1914, 1919$\\
10. Adjusted Outliergram         & $1919$             & $1919$\\
\hline
\end{tabular}}
\caption{Outliers detected by the different methods in the raw and smoothed Australian male log-mortality rate samples.\label{aus}}
\end{table}

\end{document}